\documentclass[]{aastex631}

\usepackage{comment}
\usepackage{amsmath}

\newcommand{\sampsize}{37,598}
\newcommand{\vlsr}{V_{LSR}}
\newcommand{\lbv}{$\ell-b-\vlsr$}
\newcommand{\HI}{H\,{\sc i}}

\newcommand{\hl}{}
\shorttitle{Pattern Mapping HI Maps}
\shortauthors{Craig et al.}
\graphicspath{{./}{figures/}{Figures/}}

\begin{document}

\title{A map of the outer gas disk of the Galaxy with direct distances from young stars}

\author[0000-0002-0786-7307]{Peter Craig}
\affiliation{Center for Data Intensive and Time Domain Astronomy, Michigan State University \\
East Lansing, MI 48824 USA}

\author[0000-0001-6711-8140]{Sukanya Chakrabarti}
\affiliation{Department of Physics and Astronomy, University of Alabama \\
Huntsville, Huntsville, Alabama, 35899}

\author[0000-0002-3662-3942]{Alexander R. Pettitt}
\affiliation{Department of Physics \& Astronomy, California State University Sacramento\\
6000 J Street, Sacramento, CA 95819, USA}

\author[0000-0003-3939-3297]{Robyn Sanderson}
\affiliation{Department of Physics and Astronomy, University of Pennsylvania\\
Philadelphia, PA 19104\\}

\author[0000-0002-5204-2259]{Erik Rosolowsky}
\affiliation{Department of Physics, University of Alberta, Edmonton\\
AB T6G 2E1, Canada\\}



\begin{abstract}

For more than fifty years, astronomers have mapped the neutral hydrogen gas in the Galaxy assuming kinematically derived distances. We employ the distances of nearby young stars, which trace the gas from which they formed, in longitude-latitude-velocity space to map this gas without using kinematic distances.  We denote this new method "pattern matching".  Analysis of simulated spiral galaxies indicates that our pattern matching distances are 24$\boldsymbol{\%}$ more accurate than kinematic distances for gas within 15 kpc of the Sun. \hl{The two methods provide similar agreement with parallaxes towards these masers, although the kinematic method shows a small systematic offset in the distance that is not present in the pattern matching distance.} Using parallaxes and velocities for masers, we show that this novel method, when matched with nearby Cepheids, performs well compared to kinematics.  This analysis is restricted to sources that have a reasonably good match with a member of our Cepheid sample. The distances derived here, and the associated map, have broad utility - from improving our understanding of star formation and the dynamical structure of the Galaxy, to informing 3-D dust maps.

\end{abstract}



\section{Introduction}

The distribution of gas in the MW contains a wealth of information about the structure and dynamics of our Galaxy \citep{McClureGriffiths2004,Levine2006,Koo2017,Soleretal2022} that can impact a broad range of areas in astronomy. The neutral hydrogen gas (\HI), when converted to molecular gas, can fuel continued star formation in the Galaxy \citep{Bigiel2008,Krumholzetal2009,Krumholz2012,Heyer2015,Yu2022}. The \HI{} in our Galaxy has long been used for understanding the structure of the MW disk, such as spiral arms and warps in the disk \citep{VanDeHulst1954,Kerr1957,Westerhout1957,Oort1958}. The distribution of high-mass X-ray binaries, a young Galactic population, is found to correlate with the spiral arms, i.e., with the dense concentration of gas, in the Galaxy \citep{Walteral2015}. The gas distribution is also an important input for modeling the diffuse emission in gamma rays that arises from neutral pions from the interaction of non-thermal cosmic rays with the interstellar medium \citep{Tibaldo2021,Zhang2023}.

Maps of the \HI{} in our Galaxy have traditionally been made from the Doppler shift of the 21\,cm hyperfine emission line of \HI{} by assuming a model for the velocity field \citep{VanDeHulst1954,Oort1958,Kerr1962,Henderson1982}. The assignment of these so-called kinematic distances uses an assumed rotation curve to convert velocity information into a distance estimate \citep{Levine2006,Kalberla_Dedes2008,Koo2017}. The accuracy of kinematic distances can depend heavily on the rotation curve assumed for the disk. Regions of the disk that deviate from the assumed velocity model, such as near streaming motions along spiral arms and bars or in inter-arm regions as highlighted by recent work \citep{Hunter2024}, will lead to systematically inaccurate distance estimates. This can produce misleading features in the resultant maps of the MW's gas distribution. 

\hl{Many other methods have been used over the years to model the \HI{} distribution in the MW. For instance, \cite{Soding2025} use 3D Gaussian processes in order to model correlations in the ISM, which they use to construct a 3D map of the \HI{} in the MW that recovers many known features. \cite{Mertsch2023} use a Bayesian inference technique to construct a Galactic \HI{} map for the HI4PI data. \cite{Tchernyshyov2017} and \cite{Tchernyshyov2018} build maps of \HI{} utilizing a tomographic approach using a combination of the 21 cm line emission and maps of the Galactic reddening. \cite{Murray2019} construct maps of the gas in the Small Magellanic Cloud (SMC), by comparing the gas velocities with those of stars in the SMC (which have measurable differences).}

Here, we present a novel method for assigning distances to the gas from known precise distances of the nearest young stars in longitude-latitude-velocity (\lbv) space. Employing the precise distances of Cepheid variables and other young stars (typically $\lesssim 400$ Myr), we show that we can derive model-independent distances for the gas that are somewhat more precise than kinematic distances. The precision of our method is expected to improve as the number of catalogued stars with precise distances increases. Our method for assigning distances relies on known observed correlations between the structure of the young stellar population in galaxies and  the gas \citep{Chakrabartietal2003,Quillenetal2020}. It is clear from studies of our Galaxy and external galaxies that the spiral pattern of the young stars closely traces the gas from which it is born \citep{Russeil2003,Schinnerer2017,Querejeta2021}. The correlation with the older stars, however, becomes progressively weaker as the older stars migrate from the gas from which they formed \citep{Blitz_Shu1980,Grasha2019,Sun2022,Turner2022,Peltonon2023}. 


With the advent of the \textit{Gaia} satellite, we now have velocity and distance information for many young stars throughout the Galaxy \citep{Gaia}. By finding the nearest neighbors in \lbv{} space, of the stars to the gas in the MW, we can assign precise distances from young stars to the gas. Our approach only works in the outer disk, because the inner disk suffers from dual-valued distances (i.e. the "distance ambiguity") \citep{Levine2006}. This limitation also applies to kinematic distances, where rotation curve models indicate that the distances are double-valued in the inner disk. An additional limitation of both methods is that at longitudes near $0^\circ$ and $180^{\circ}$, the line-of-sight velocities are close to zero and are distance-independent, so velocity information cannot be used to assign distances.

Our young stellar sample, with a median age of 377 Myr, is constructed from recent work on Cepheids and masers in the MW \citep{Skowron2019,Dekany2019,Reid2019,Drimmel2023}, combined with Gaia DR3 data \citep{Gaia,GaiaDR3}. In total, we have \sampsize{} stars distributed throughout the disk, mostly between $R = 7$ kpc and $R = 31$ kpc. The sample is required to have $R > 7$ kpc because gas associated with stars at smaller radii will be excluded from our maps due to the aforementioned distance ambiguity. While we do not impose a specific outer distance limit in our sample, all sources are required to have distance and radial velocity measurements for our method. This leads to a sample that extends from $7-35.5$ kpc.


In our sample of \sampsize{} stars, there are 3,502 Type I Cepheid variables and 199 masers \citep{Reid2019}. The remaining stars are sources from \textit{Gaia} with parallax-derived distances. \textit{Gaia} data is added as needed to supplement the sample. The detailed construction of the young stellar sample, including the choice of age threshold and uncertainty cuts, is described in Appendix \ref{app:stars}. \hl{We first built maps using \HI{} data from the Leiden-Argentine-Bonn survey (LAB) \citep{Kalberla2005}, matching the data set used in \citep{Levine2006} for comparison purposes. Additionally, maps are constructed utilizing the newer HI4PI data \citep{Bekhti2016}, where we generate both pattern matching and kinematic maps.}


\begin{figure}
\begin{center}
\includegraphics[width=\columnwidth]{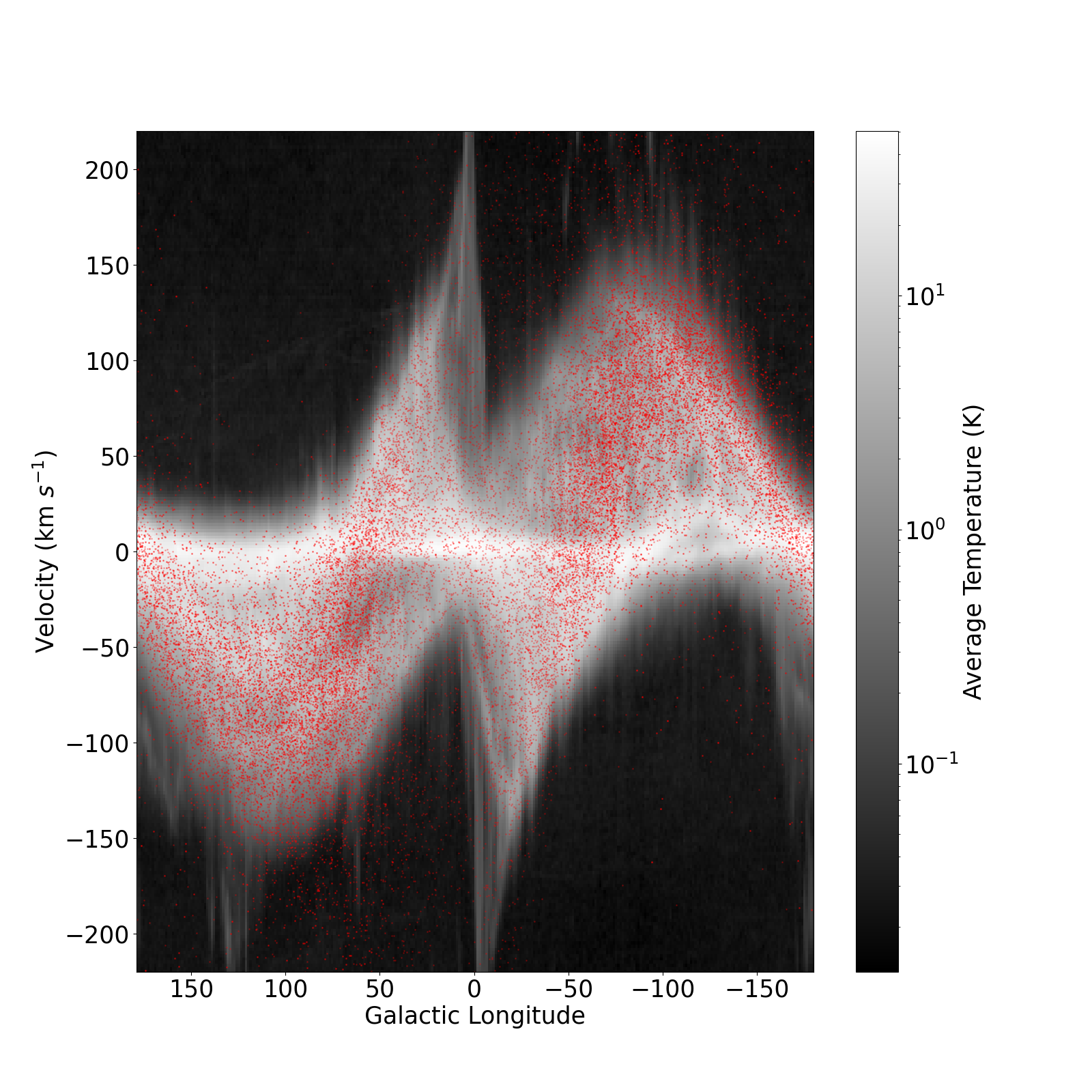}
\caption{\HI{} brightness temperatures in the $\ell$ vs. $\vlsr{}$ plane. In each bin, the \hl{brightness} temperature has been averaged over Galactic latitude. The red points display the positions of the stars in our stellar sample. Our sample of young stars samples all the major features in the gas in this plane and contains valuable distance information that can be applied to the underlying gas. 
} \label{fig:Lab_Cepheid_lvlsr}
\end{center}
\end{figure}

We remove gas at high latitudes, $|\mathrm{b}| >$ 30 degrees, that is not associated with the MW disk. We follow the procedure described in \cite{Levine2006} to remove small features in the \hl{brightness} temperature distribution, with details provided in Appendix \ref{app:param}. Figure \ref{fig:Lab_Cepheid_lvlsr} depicts the brightness temperatures of the \HI{} gas in velocity-longitude space along with our stellar sample. These brightness temperatures are averaged over latitude. As is clear, \HI{} gas closely traces our young stellar sample. The significance of this correlation can be determined by evaluating the likelihood that the gas and stars are drawn from the same distribution. This likelihood can be contrasted with the observed young stellar sample that we employ here, and a randomly generated sample uniformly distributed throughout the disk. The likelihood of the former distribution is much larger by more than 100 orders of magnitude. As the young stars are born recently from the gas, it is not surprising that they are a better tracer of the gas. Additional details about this analysis can be found in Appendix \ref{app:likelihoods}.

\section{Methods}

Given that our algorithm matches sky positions and velocities of gas to nearby stars, and subsequently assigns the known distance of nearby stars to the gas, we call it ``pattern matching". We do this by finding the closest stars in the \lbv{} data cube, which is determined using Equation \ref{eqn:distance_metric}. This defines a distance in \lbv{} space, which we will refer to as $s_{LBV}$ to avoid confusion with heliocentric distances. \hl{We note that this selected metric does not include the uncertainties on the stellar velocity and distance measurements, or information on the stellar age. These parameters could also be considered to give more weight to stars that are expected to be better suited for tracing the underlying gas distribution. These are left out of Equation \ref{eqn:distance_metric} because of the way that our stellar sample is constructed, which generally tends to have stars in particular regions of the galaxy with similar uncertainties. In regions where the stellar sample includes stars with larger uncertainties or ages, then the number of suitable stars with good agreement in \lbv{} space is small, and stars with a good match in \lbv{} space are more likely to be good tracers of the gas than a younger star (or one with smaller uncertainties) with a worse match in \lbv{} space. In effect, we do not include the uncertainty information in the metric here because the stars in each region of the disk have already been selected as the best stars to match with the gas in that region.}

\begin{equation}\label{eqn:distance_metric}
s_{LBV} = \sqrt{\frac{ (\ell_1 - \ell_2)^2 }{1^{\circ 2}} + \frac{(b_1 - b_2)^2}{{1^{\circ 2}}} + \frac{(v_1 - v_2)^2}{(1 \mathrm{km\ s^{-1}})^2}}
\end{equation}

For each gas measurement, we use up to $N_{\mathrm{star}}$ nearby stars and take a weighted average of their distances weighted by the inverse of their $s_{LBV}$ distance. For the LAB data, we select $N_{\mathrm{star}} = 3$. We have explored a range of values for $N_{\mathrm{star}}$, and the resulting maps are similar as long as $N_{\mathrm{star}} > 1$, as described in Appendix \ref{app:param}. A weighted average was selected over a median to mitigate problems in regions with sparse stellar samples. In this case, there may only be one star with a small $s_{LBV}$ value and the median distance assigned can correspond to stars with substantially larger $s_{LBV}$. The weighted average will instead primarily use the distance of the one low $s_{LBV}$ source. We have tested several metrics for $s_{LBV}$, and find that they provide similar surface density maps, as discussed in Appendix \ref{app:param}.

To carry out our pattern matching technique, we create a regular grid in the $\ell$ vs. $\vlsr{}$ plane, then identify the stars that are contained in each bin. For a given gas measurement, we calculate $s_{LBV}$ for all stars contained in the bin containing the $\ell$ and $\vlsr{}$ values for the gas, and the eight adjacent bins. This is a linear nearest neighbor search over this subset of the stellar population, which is guaranteed to find any stars within one bin width. The bin size for this grid is set based on the number of stars in the stellar sample. For samples with a large number of stars ($\sim$ 10$^{5}$), as can occur in simulations, this is set to 2, ensuring that any stars within 2 degrees in longitude and 2 km s$^{-1}$ in $\vlsr{}$ of the gas are included. If the stellar sample contains on the order of $10^4$ stars, such as our young stellar sample, this value is set to 3, ensuring that gas measurements in regions with a less dense stellar population still have an adequate number of reasonable stars to match with. Stars outside this radius are unlikely to be accurate distance tracers and are therefore excluded. The distance associated with this point in \lbv{} is then calculated as a weighted average across the best $N_{\mathrm{star}}$ stars. In case the number of nearby stars is 0, the gas will not be included in the final maps as viable distances cannot be assigned.

Next, the gas positions in \lbv{} space, combined with distance measurements, are transformed into Galactocentric cylindrical coordinates. For LAB HI data, this results in an irregular grid in these coordinates. We then convert the brightness temperatures into mass densities at each grid cell. \hl{This conversion is done following the method in \citep{Kerr1968}, assuming a constant spin temperature of $T_{s} = 255$ K. A benefit of this approach is that it is insensitive to the properties of the assumed grid, so our results do not vary significantly with changes in the assumed grid structure and scale.} In order to produce maps of the surface density, it is convenient to have a regular grid in these coordinates. We create a regular grid in R, $\phi$, and Z, with R ranging from 8.5 kpc to 30 kpc with 100 grid values in radius. In $\phi$, we cover $-\pi$ to $\pi$ with 350 samples. In the Z direction, we use bounds at -20 and 20 kpc with 141 samples. This is our standard grid, and is used in all the analysis in this paper. Finally, we compute the density in each grid cell as the average density across the gas measurements contained within the grid cell.

To determine the accuracy of kinematic and pattern matching distances, we have analyzed various hydrodynamical simulations. Our fiducial simulations are intended to be simple, and consist of a gas disk evolving in a set of analytic, time-static potentials that represent the stellar and dark components of a MW-like galaxy and a rigidly rotating spiral pattern \citep{Pettitt20}. These simulations have been re-scaled by a factor of 2.5, such that the new stellar population can trace out a region similar in size to the region of interest in the MW. These simulations were conducted with the \textsc{Gasoline2} \citep{Wadsley2017} smoothed particle hydrodynamics code and include gas self-gravity, cooling, star formation, and supernova feedback. The resolution of the gas and young star particles is 2000$M_\odot$ and the gravitational softening length is set to 50 parsec. The main simulation used here includes two spiral arms, and will be referred to as SP2. Other simulations that we have analyzed include a time-dependent dark matter halo with a dwarf galaxy perturber, along with prescriptions for converting gas to stars \citep{Chakrabarti_Blitz2009, Chakrabarti_Blitz2011,Chakrabartietal2019} are discussed in Appendix \ref{app:sims}. 



The simulated stellar samples are chosen to only contain star particles that formed during the simulations, approximating our real stellar sample. All included particles are younger than 400 Myr, with a nearly uniform age distribution from 0 to 400 Myr. The resulting sample contains 59,690 star particles. Gaussian uncertainties on the distances and $\vlsr{}$ values are then added, set to be 10$\%$ uncertainties on distance and $2$ km s$^{-1}$ on the velocities. Finally, we randomly select \sampsize{} sources to match the size of the MW stellar sample. Then the simulated sample approximately resembles the true stellar sample, although with a lower average age of 200 Myr. The resulting maps are not strongly sensitive to the selected stellar age, and maps generated that include an older stellar population closely resemble the fiducial case.

We consider several rotation curves to produce kinematic distance estimates in the simulations. The simplest model uses a flat rotation curve for the outer MW disk, and serves as the primary model for our rotation curves. This is designed to compare to  the approach used in building the LAB \HI{} map presented in \cite{Levine2006}. Our kinematic model is primarily used for comparison against our pattern matching techniques, so we assume circular orbits to provide a baseline to compare against. We have considered including radial components for the gas kinematics as done in \cite{Levine2006}, but the differences between models are small enough that they will not impact the main conclusions of this paper. Otherwise, we follow the methods outlined in \cite{Levine2006}. We adopt the same grid used for our pattern matching techniques described above. Densities are assigned to each grid cell by taking the average density of the particles contained within each cell. The remaining analysis can be performed using the same techniques as used for the observed data.

\section{Results}

\begin{figure*}
\begin{center}
\includegraphics[width=1.0\textwidth]{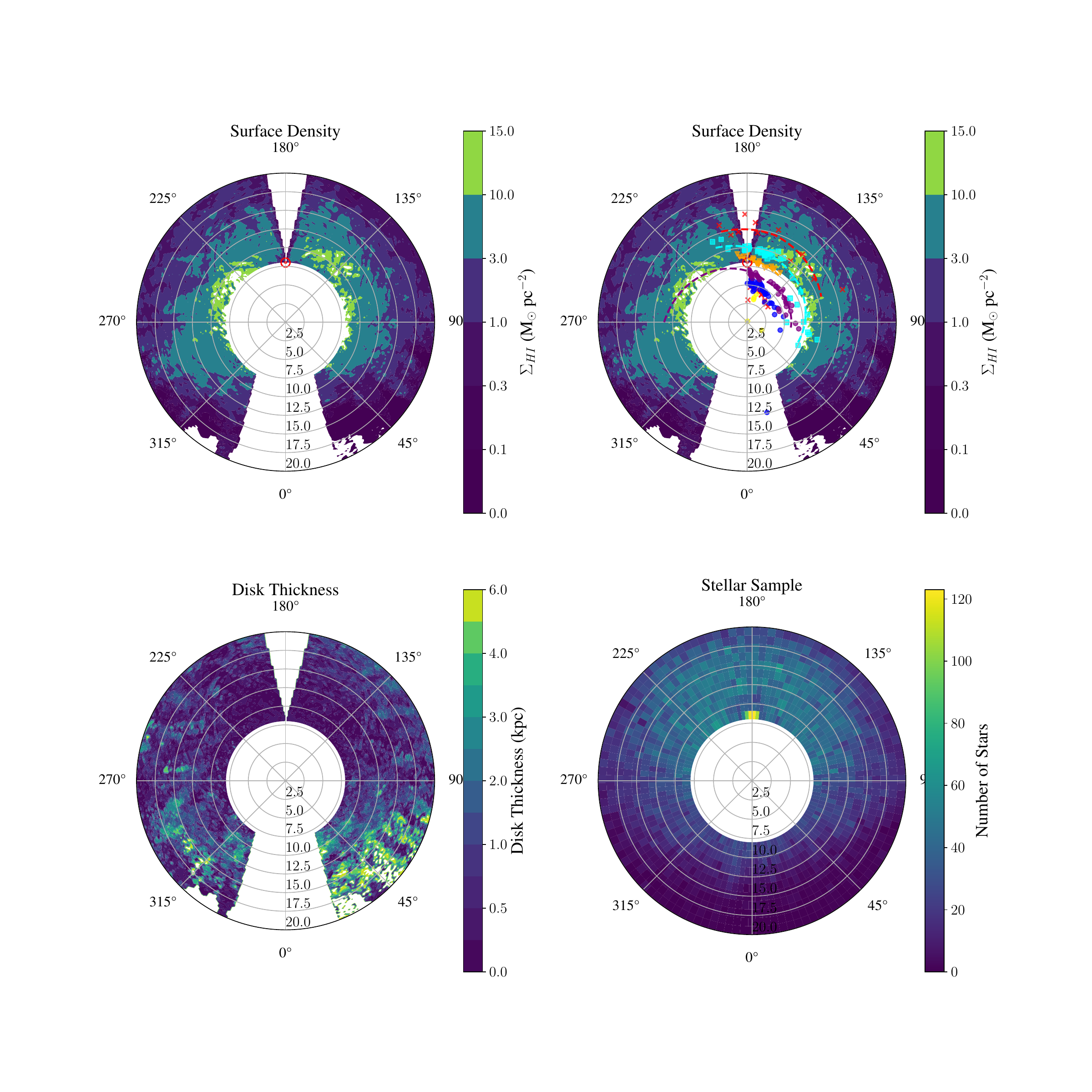}
\caption{The top left panel shows the surface density map derived using our pattern matching technique. \hl{In the top right panel, we display the Maser locations, color-coded by spiral arm, plotted along with the spiral arms from \cite{Reid2019} on top of the pattern matching surface density}. The arms shown are, in order of increasing radius, the Local arm (orange), the Perseus arm (cyan), the Outer arm (red), and the Sagittarius-Carina arm (purple). We only show spiral arms that are fit to masers in the outer disk and match high density features in our surface density map. Gaps in this map in the outer regions of the disk are a result of the lack of stars in those regions with suitable measurements, and will improve in future iterations using stellar samples from future \textit{Gaia} data releases and from \textit{Roman}. \hl{In the bottom left, we show the disk thickness map derived utilizing the pattern matching distances}. The bottom right panel displays a 2-D histogram with the number of stars in the same coordinate system, where we can see that the stellar sample falls off in the outer disk. Regions with no stars in this sample will also have no gas assigned, and the well sampled regions will produce more accurate gas distances.} \label{fig:HIMap}
\end{center}
\end{figure*}

\begin{figure*}
\begin{center}
\includegraphics[width=1.0\textwidth]{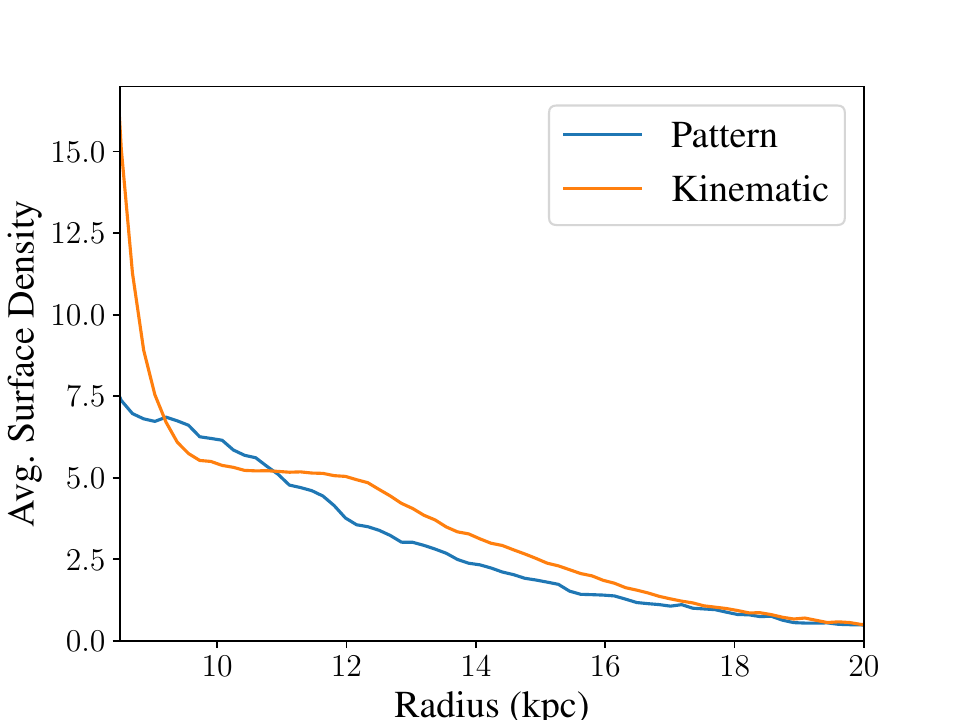}
\caption{Average surface density profiles for our pattern matching and kinematic maps. Here we see some disagreement at small radii, followed by a somewhat lower average density from pattern matching at larger radii.} \label{fig:RadProf}
\end{center}
\end{figure*}

\begin{figure*}
\begin{center}
\includegraphics[width=1.0\textwidth]{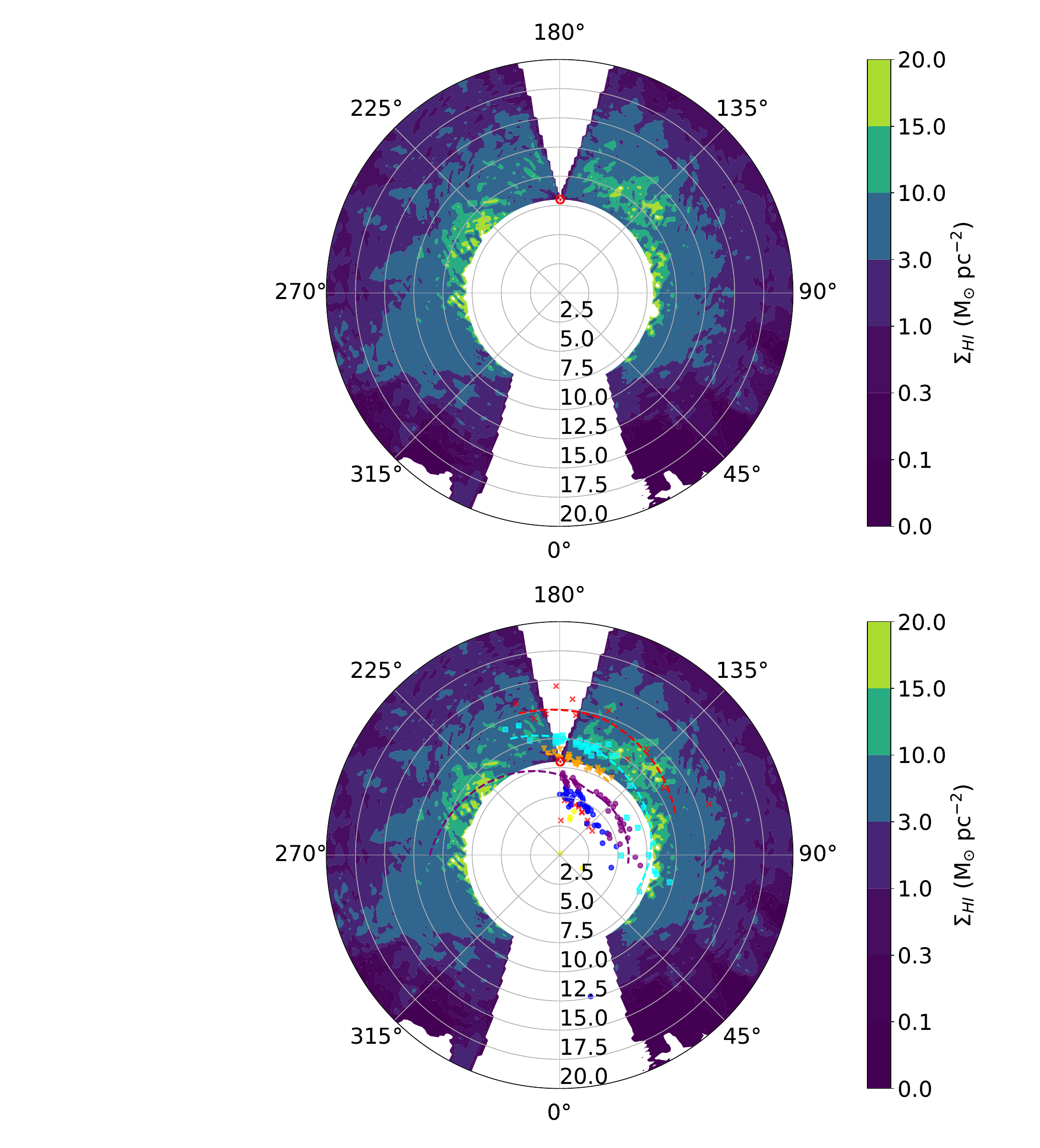}
\caption{\hl{The top panel shows the surface density map for the HI4PI data set, derived using our pattern matching technique. The bottom panel displays the same map, but with the Maser locations, color-coded by spiral arm, plotted along with the spiral arms from \cite{Reid2019}. The arms shown are, in order of increasing radius, the Local arm (orange), the Perseus arm (cyan), the Outer arm (red), and the Sagittarius-Carina arm (purple). We only show spiral arms that are fit to masers in the outer disk and match high density features in our surface density map. Gaps in this map in the outer regions of the disk are a result of the lack of stars in those regions with suitable measurements and will improve in future iterations using stellar samples from future \textit{Gaia} data releases and from \textit{Roman}.}} \label{fig:HI4PI_Map}
\end{center}
\end{figure*}

\begin{figure*}
\begin{center}
\includegraphics[width=1.0\textwidth]{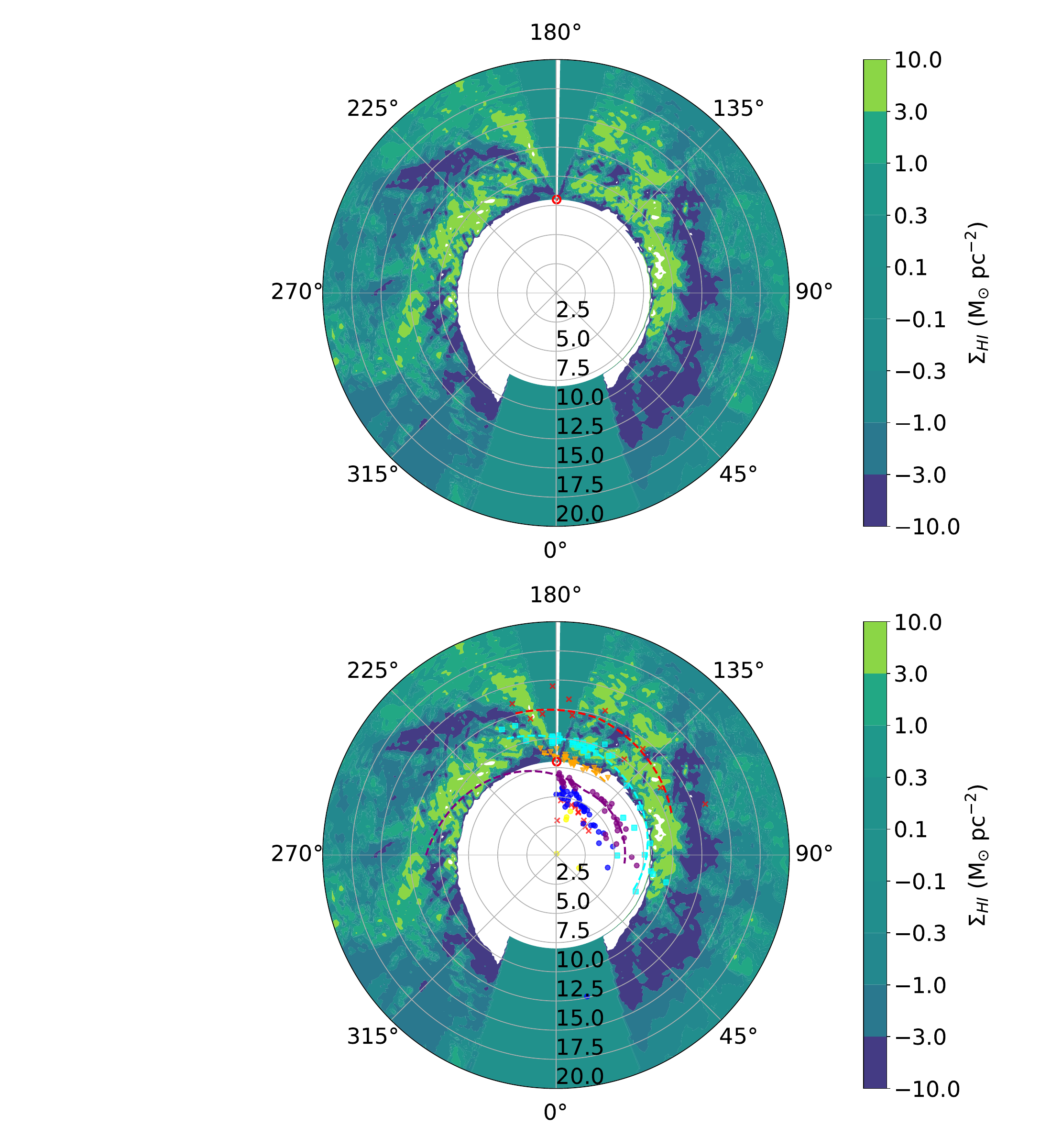}
\caption{\hl{Residuals between the kinematic and pattern matching techniques (specifically pattern matching - kinematics) for the HI4PI data. The residuals in some places are fairly significant, though variability on small scales is likely unreliable. The larger scale differences are most likely due to disagreements about the precise locations of high density regions in these maps, yielding large positive and negative residuals around the locations of spiral arms. Many of the significant residuals trace out the locations of the spiral arms derived in \citep{Reid2019}.}} \label{fig:HI4PI_Residuals}
\end{center}
\end{figure*}

Figure \ref{fig:HIMap} shows the pattern matching surface density map for the \HI{} disk of the MW. The bottom left panel of this figure displays the distribution of the stars in our sample. There are some gaps, especially at large heliocentric distances, which correspond to the gaps in our derived \HI{} map. While our derived maps extend out to 30 kpc, the results shown here are truncated at a Galactocentric radius of 20 kpc. Our map becomes jagged and incomplete in some regions, especially on the far side of the MW where Galactic extinction prevents the creation of a suitable stellar sample. This is a current limitation of our methods, but should improve over time with better data availability, especially future \textit{Gaia} data releases, and \textit{Roman} data in the future.

Figure \ref{fig:RadProf} displays the radial density profile for the gas, derived using our pattern matching map and the kinematic map. Here we see that the profiles are similar at most radii, with the pattern matching map producing somewhat smaller densities at low radii, and similar densities elsewhere. At larger radii between 20 and 30 kpc, which are not shown here due to the degrading quality of the stellar sample, the pattern matching density becomes larger. \hl{Gas in this outer region may have systematically underestimated distances, since most nearby stars will be closer than the gas measurement when the stellar sample tapers off. The result of the fading stellar sample in some places in the disk is that most of the gas beyond the extent of the stellar sample will be assigned the distance at the edge of the stellar sample coverage. As a result, some regions of increased density can be formed inaccurately due to many gas measurements being assigned the same (or quite similar) distances. For this reason, the density at the edge of the derived surface density map is considered to have some systematic bias towards higher densities. This only applies to the edge bins of the disk; inwards of these bins, this stellar sample is sufficiently populated that this is not a significant problem.}

\hl{The maps showing the LAB data are useful for making comparisons with previous work. We also have generated pattern matching maps using the HI4PI data, which can be seen in Figure \ref{fig:HI4PI_Map}.  The large scale features seen in this map are similar to those observed in the LAB data, with some differences in the small scale structure. There remain high density regions in these maps that are well aligned with the spiral arms indicated by the masers.}

\hl{In order to compare between the kinematic and pattern matching techniques, we compute residuals between the two for the HI4PI data. This is displayed in Figure \ref{fig:HI4PI_Residuals}, where we do see some large residuals between the two cases. These residuals typically arise from disagreements between the two methods about the distances to high density regions, as well as the radial width of these regions. The result is a significant positive residual that is adjacent to a significant negative residual. This kind of disagreement can be explained by an offset in the rotation curves used for the kinematics compared to the true gas kinematics. This, coupled with differences in the distance assignments leading to different widths of the high density regions, drives the residual features in this map. In the outer disk, the two maps approximately agree with each other. For example, there is a large positive residual at 275 degrees with a radius of 12.5 kpc, with a significant negative residual at a radius closer to 10 kpc. This arises from a distance mismatch between the two methods on the order of 1-2 kpc. If the pattern matching distance to the higher density gas is reduced slightly, then the residuals would be much smaller.} 

Figure \ref{fig:Fourier} shows the amplitudes of the $m=1$ through $m=4$ Fourier modes, normalized to the axisymmetric or $m=0$ mode, for the maps of the \HI{} data. The power in the low-order Fourier modes for both maps is similar at lower radii. Interestingly, the pattern matching map has significantly more power in the $m=1$ mode at larger radii. The $m=3$ and $m=4$ modes have less power in the pattern matching map compared to the kinematic map.

There are several high surface density regions manifest in this map, especially at smaller radii, which are tracers of spiral structures in the MW disk. The low radii areas have better stellar data available, and therefore are expected to have the highest quality \HI{} map. These high density regions are well aligned with the maps of spiral structure in the MW disk generated based on maser data, and our map generally agrees with the properties of the Local, Perseus, and Sagittarius-Carina arms derived from masers. These three arms are marked in Figure \ref{fig:HIMap} as identified from the maser data of \cite{Reid2019}. Overall, the resulting map is relatively flocculent compared to previous \HI{} maps \citep{Levine2006}.

\hl{We measure the flocculence of the spiral arms based on the ratios of Fourier amplitudes for the maps. This is done utilizing the $f_3$ parameter following \citep{Yu2020}, with $f_3$ defined in equation \ref{eqn:f3} where the $A_{n}(r_i)$ are the Fourier amplitudes of the nth mode at radius r. Larger values of $f_3$ tend to indicate that the spiral arms are more flocculent, as flocculent arms will tend to appear with higher-order frequency structures. We find that the Fourier amplitudes are consistent with the new pattern matching maps being more flocculent than the kinematic maps. In particular, we recover amplitude ratios of 0.24 for the kinematic map and 0.38 for the pattern matching map.}

\begin{equation}
f_3 = \frac{1}{N} \sum^{N}_{i=1}\sqrt{\frac{A_3^2(r_i)}{A_2^2(r_i) + A_3^2(r_i) + A_4^2(r_i)}}
\end{equation}\label{eqn:f3}

We do not see an extended Outer Arm (i.e., with a long azimuthal extent towards $\phi \sim 270^{\circ}$ as in \cite{Levine2006}) in our \HI{} map. There is a region of increased surface density in the \HI{} at $R \sim 10$ kpc near azimuth $\sim 135$ degrees, but this is a relatively small clump that lacks the large azimuthal extension of the Outer Arm as shown in \cite{Levine2006}.  The long extent of the Outer Arm that is shown in \cite{Levine2006}, extending from $\sim 180^{\circ}$ to $\sim 270^{\circ}$, is not seen in the maser data of \cite{Reid2019}. An arm-like feature is identified in the \cite{Reid2019} data, but over a smaller extent, close to $\phi \sim 180^{\circ}$. Moreover, the appearance of this extended Outer Arm in the \cite{Levine2006} map is quite different from the other features in the \cite{Levine2006} map that are more flocculent in nature, as expected in the gas \cite{Chakrabartietal2003}. Taken together (the lack of a comparable feature in the maser data as well as in our map) and distinct appearance of this arm compared to others in \cite{Levine2006}, we are led to suggest that the Outer Arm (or at least the extent of it) may be spuriously identified in prior \HI{} maps. Indeed, the Outer Arm is rarely seen in the third quadrant in stellar tracers \citep{Uppal2023}, so it is not surprising that we do not recover it using our technique.


The thickness of the gas disk has been computed following the calculations in \cite{Levine2006}. The resulting disk thickness map is shown in Figure \ref{fig:HIMap}. The disk thickness is increasing at larger radii, corresponding to locations in the disk with lower surface densities. \hl{We note that at large distances from the Sun, the distances to the stellar sample are less accurate, which might contribute to the disk thickness as a result of stars with poorly constrained heights off the disk. On the near side of the disk, the distance errors in our sample are not sufficient to explain the increase in the disk thickness observed in the pattern matching maps, although it is possible that it contributes to the increased disk thickness at large heliocentric distances. Such an effect could explain why the disk thickness estimates are increased on the far side of the disk, but even considering this, the pattern matching maps still indicate that the thickness on average increases at larger radii.} This indicates that the thickness of the disk is anti-correlated with the surface density in the disk, as found in earlier work \citep{Henderson1982,Levine2006,Soleretal2022}. This anti-correlation is a prediction of gravitational stability analysis of gaseous discs.

\subsection{Simulated \HI{} Maps}

\begin{figure*}
\begin{center}

\includegraphics[width=\textwidth]{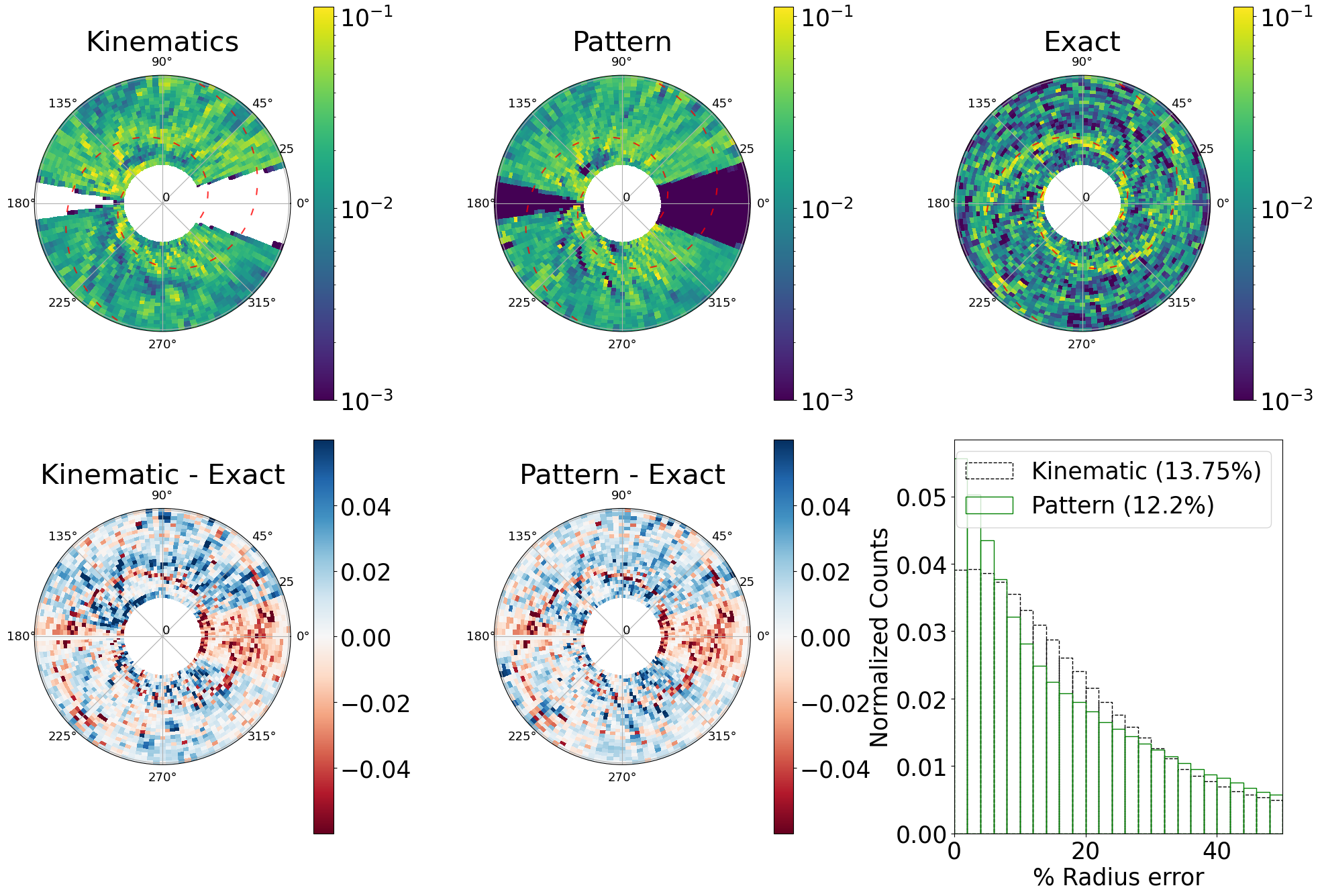}

\caption{Surface density maps from the SP2 simulation, including both the derived maps from kinematics and pattern matching as well as the exact gas distribution. Dashed red lines overplotted on these maps display the positions of the two spiral arms. The bottom left and middle panels show residuals (derived map minus true map), which indicate a better match from the pattern matching case. The bottom right panel presents histograms of the percentage accuracy of the measured Galactocentric radii from each method. These show the distribution of the percentage errors for the gas particles for the kinematics and pattern matching techniques, where the pattern matching is more likely to measure accurate radii. The figure legend in this panel displays the median percentage accuracy for each method.} \label{fig:sim_results}
\end{center}
\end{figure*}

By comparing the gas surface densities determined using kinematic and pattern matching distances in SP2, we can assess their relative performance. The surface density maps derived by pattern matching are generally more accurate than those generated using kinematic distances, as seen in Figure \ref{fig:sim_results}. Across the full disk of the simulations, we find that the median distance error is $20.0\%$ for the kinematics, and $16.2\%$ for the pattern matching. In order to better compare with observational accuracy estimates from the maser sample, we also consider the accuracy within 15 kpc of the Sun, which yields a median distance accuracy of $23.6\%$ and $14.9\%$ for the kinematics and pattern matching, respectively. As seen in Figure \ref{fig:sim_results}, the accuracy of the derived Galactocentric radii is also better for the pattern matching technique. 

Without the measurement uncertainties added to the simulated data, we instead find that the distance accuracy is $19.5\%$ for kinematics and $14.4\%$ for pattern matching. Within 15 kpc this becomes $23.6\%$ for kinematics and $13.2\%$ for pattern matching, \hl{representing a significantly better distance accuracy when derived through the pattern matching technique.} Increasing the number of stars to the full number of young star particles in the simulation leads to a pattern matching accuracy of $14.3\%$. Within 15 kpc, this gives an accuracy of $13.2\%$.

To quantify the improvement in the surface density maps, we calculate a residual by subtracting the true surface density in the simulation from the map generated by each method. Then, we can compute a sum of the square residuals, providing a measurement of the quality of a map. We divide this by the average surface density of the kinematic map, which in this case is to 0.023 M$_{\odot}$ pc$^{-2}$. Across multiple simulations, the pattern matching residuals are consistently lower. For the SP2 simulation, this value yields 56.5 for the pattern matching map and 62.0 for the kinematic map, both in units of $\rm M_{\odot} pc^{-2}$, indicating that the pattern matching map is more accurate than the kinematic map.

\begin{figure*}
\begin{center}
\includegraphics[width=1.0\textwidth]{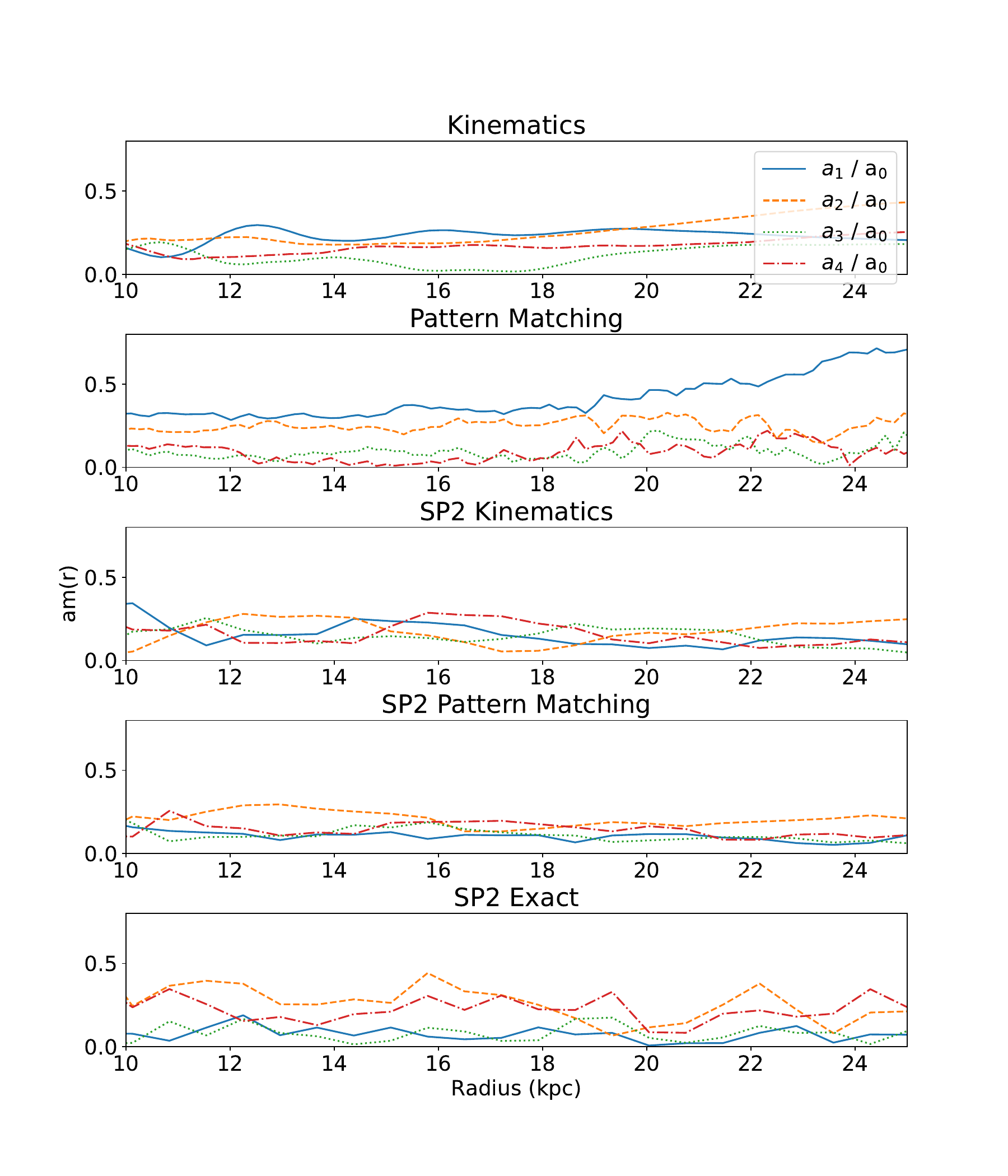}
\caption{The top panel displays the Fourier amplitudes for the kinematic map of the LAB \HI{} data, with modes 1 through 4 shown normalized against the m=0 mode. The bottom panel shows the same amplitudes for the pattern matching case. The following three panels show the Fourier amplitudes derived for kinematic, pattern matching, and true maps of the SP2 simulation.} \label{fig:Fourier}
\end{center}
\end{figure*}

We have also compared the power in the Fourier modes for maps constructed using both techniques, following a similar analysis to that in \cite{Chakrabarti_Blitz2009,Chakrabarti_Blitz2011}. The results for the SP2 simulation (along with the LAB HI results) are displayed in Figure \ref{fig:Fourier}. Here we see that the pattern matching more accurately approximates the true map in the $m=1$ and $m=2$ modes compared to the kinematic map. To quantify this, we compute the average percentage error in the Fourier modes for each map. In the SP2 simulation, the $m=1$ mode has lower errors by a factor of 2.2 for the pattern matching method, and the errors are lower by a factor of 4.6 in the $m=2$ mode. For $m=3$ and $m=4$, the kinematic map performs slightly better, with errors lower by a factor of 1.4 for both modes.



As an additional check, we test our algorithm against the observations directly, using our pattern matching method to compare our inferred distances against maser parallaxes from \cite{Reid2019}. First, a new stellar sample is constructed that is limited to the local Cepheids. Next, distances from these young stars are assigned to the masers using our pattern matching method. Kinematic distances are also assigned for comparison. We compute the difference between the parallax distances and those recovered from each method to define a distance residual. The median and average values of these samples are comparable (within 0.3 kpc of each other, but slightly larger for the pattern matching). \hl{The kinematic method has a tendency to systematically overestimate the maser distance compared to the kinematics, however}. This is restricted to 53 masers that are in the outer disk and have viable distances through both methods. Including inner disk masers produces much better pattern matching performance compared to kinematics, although this may be due to the handling of the distance ambiguity.


The masers were also tested using our pattern matching technique, this time without the masers included in the stellar sample (otherwise the pattern matching just outputs the maser parallax distances directly). This case yields a worse match from pattern matching. This is most likely caused by \textit{Gaia} sources that are relatively old, used to increase the density of the stellar sample at larger distances. This mostly impacts masers at larger radii, those located closer to the Sun tend to have a better or comparable match from pattern matching. Over time this will improve as the stellar sample is updated with younger stars and more accurate distances and velocities. As a final check of our methods, we randomly select 200 Cepheids from our sample and predict those distances using pattern matching and kinematics. The results vary depending on the chosen sample, but typically the pattern matching performs better here, with average absolute residuals lower by 10$\%$ or more. This is not particularly surprising, as the Cepheids are good tracers of their own kinematics.

\section{Summary}

Tests on this novel pattern matching algorithm using hydrodynamical simulations and comparison to maser parallaxes indicate that this technique provides accurate distances to the gas in the MW without requiring assumptions inherent to kinematic distances. This allows us to build an accurate model-independent surface density map of the neutral hydrogen in the MW. In principle, these distances can be used to supplement 3-D dust maps that are presently restricted to the solar neighborhood \citep{Greenetal2019}. In conjunction with recent direct acceleration measurements, \cite{Chakrabartietal2020,Chakrabartietal2021,Chakrabartietal2022} that give the most precise probe of the mass distribution (the stars and the dark matter) in the Galaxy, these more direct distances could be used along with direct acceleration measurements to construct an improved potential of the Galaxy. The differences between our \HI{} map and previous maps are likely to have important implications for understanding large-scale disturbances in the gas disk \citep{Levine2006,Alves2020} and how they are impacted by interactions between the MW and dwarf galaxies \citep{Weinberg_Blitz2006,Chakrabarti_Blitz2009,Chakrabartietal2019,Haines2019}. In the future, this novel distance method can be improved with more complete stellar samples, both with future \textit{Gaia} data releases as well as future \textit{Roman} data \citep{Paladini2023}, leading to increasingly accurate surface density maps that will eventually outperform kinematic maps throughout the disk. 

\section*{Acknowledgements}

This work has made use of data from the European Space Agency (ESA) mission
{\it Gaia} (\url{https://www.cosmos.esa.int/gaia}), processed by the {\it Gaia}
Data Processing and Analysis Consortium (DPAC,
\url{https://www.cosmos.esa.int/web/gaia/dpac/consortium}). Funding for the DPAC
has been provided by national institutions, in particular the institutions
participating in the {\it Gaia} Multilateral Agreement. This work has also made significant use of Python 3.11.8, and the \textsc{numpy}, \textsc{scipy}, and \textsc{matplotlib} packages.

SC acknowledges support of the National Science Foundation (NSF) Grant AAG 2009828. ER acknowledges the support of the Natural Sciences and Engineering Research Council of Canada (NSERC), funding reference number RGPIN-2022-03499. PC acknowledges the support of NASA grants 80NSSC23K0497 and 80NSSC23K1247.  SC thanks Enrico Ramirez-Ruiz and Noemie Globus for helpful discussions.  \\ 

SC dedicates this paper to the memory of Frank H. Shu. 

\bibliography{HIbib}{}
\bibliographystyle{aasjournal}

\appendix{}

\section{Stellar Sample}\label{app:stars}

We developed a method for constructing a stellar sample that is designed to balance the quality of the included stars, and the requirement for having stars distributed throughout the disk. In order to accomplish this, we begin by selecting a sample of high quality candidates for the stellar sample. The first step is the construction of a sample of classical Cepheids from \cite{Skowron2019,Drimmel2023,Dekany2019}, which have accurate distances available and are expected to be reasonably young stars. \hl{Typically, these stars are less than 200 Myr old and have distances accurate to the 0.2 kpc level, often corresponding to uncertainties on the order of $5 \%$ or less. Only $5 \%$ of the Cepheids in the sample have estimated ages above 300 Myr.} Secondly, we construct a sample from Gaia DR3 that have age estimates under 210 Myr, radial velocity uncertainties less than 150 m s$^{-1}$, and parallax over error of 95 or better.

This gives us a sample of several thousand sources, but only covers the local region of the MW disk, and will only be able to build an HI map limited to this region. In order to extend to larger distances, we relax the constraints on age, radial velocity error and parallax over error in our Gaia query. We aimed to select the best candidate stars in each part of the MW disk. We construct a 3D grid in Galactocentric cylindrical coordinates. The bins have a width of 1 kpc in R, 2 kpc in Z, and 2 degrees in $\phi$. A minimum number of stars, set to 5 for our main sample, is desired in each bin. For a given bin, if the Cepheids, masers and young Gaia sources have at least 5 sources in that bin, we do not add additional sources. If there are less than 5, we will add sources with relaxed constraints on the age and uncertainties as necessary to achieve 5 stars in that bin.

\begin{figure*}
\begin{center}
\includegraphics[width=\textwidth]{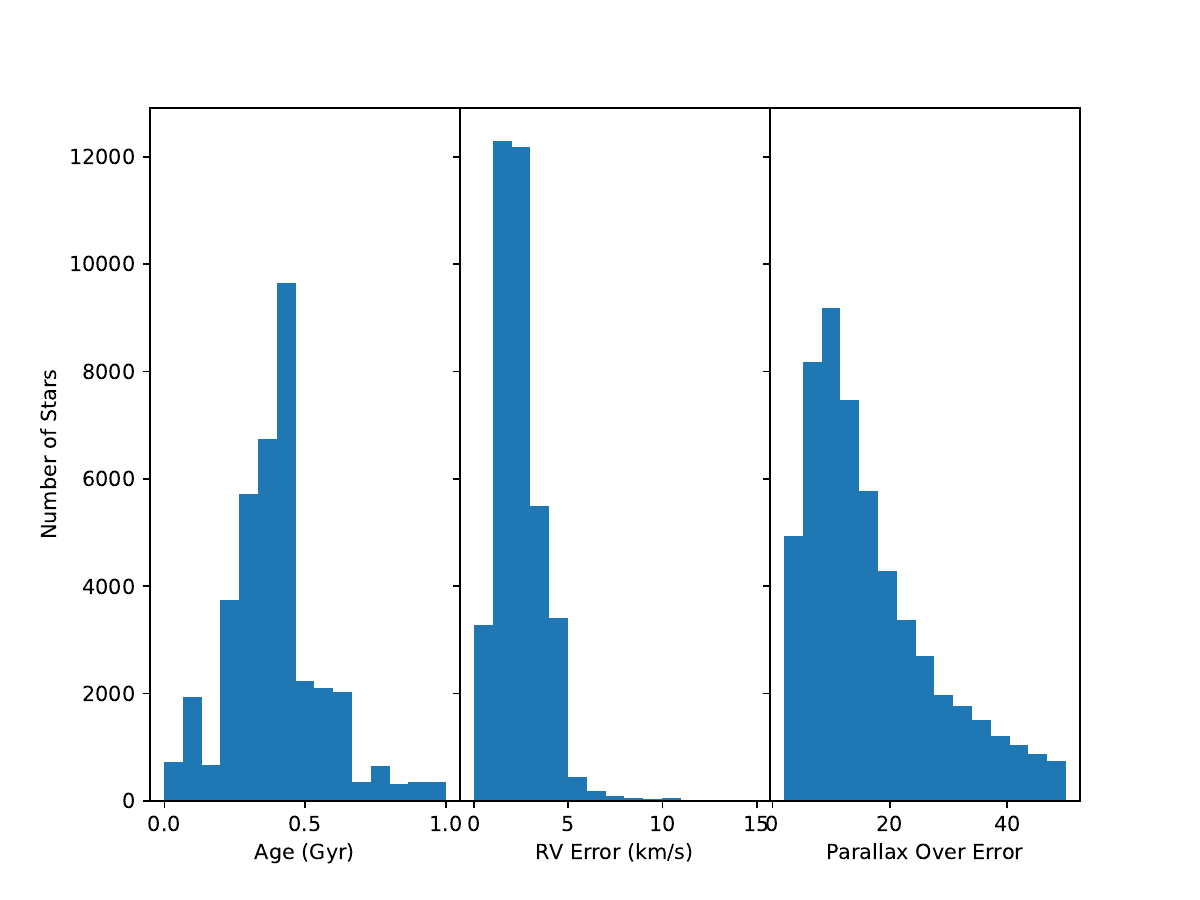}
\caption{\hl{Histograms displaying the age, RV uncertainty, and the parallax over error across our stellar sample. The histogram of stellar ages, on the left, includes the age distribution for our entire sample. In the middle panel, displaying the rv uncertainties, the tail of the distribution continues to larger radial velocity uncertainties, which has been truncated for figure purposes. There are only 16 sources out of our sample of \sampsize{} stars beyond this point, however. Our parallax over error distribution extends upwards to past 100, although the bulk of the systems are between 2 and 50, which is shown here. The displayed distribution in the right panel shows the least accurate parallaxes in our sample.}} \label{fig:stellar_params}
\end{center}
\end{figure*}

To add sources, we run a series of Gaia queries with age cuts at 400 Myr, 500 Myr and 1 Gyr, \hl{with relaxed cuts on the uncertainties at the larger age cuts}. For each query, we identify all of the stars that lie in bins that are lacking sources. Then we iterate through these bins, and add sources as necessary in order of increasing uncertainties, determined by adding the radial velocity uncertainty and 10 times the distance uncertainty in quadrature. The increased weight on the distance uncertainties is intended to eliminate sources that have highly uncertain distances, which ideally will not be used to to assign gas distances. Once a bin has the desired 5 sources, we stop adding additional stars to avoid including sources with uncertainties that are larger than necessary. \hl{This procedure is repeated for each Gaia query, in order of increasing age cut, such that bins that may be filled with the youngest stars possible (that have acceptable measurement uncertainties). The final distributions for the Age, radial velocity uncertainty, and parallax over error across this sample are shown in Figure \ref{fig:stellar_params}.}

\section{Parameter Determination and Data Reduction}\label{app:param}

The LAB \HI{} data is prepared following a similar analysis to the one presented in \cite{Levine2006}. We begin with the LAB HI brightness temperature in a regular grid in \lbv{} space. Following \cite{Levine2006}, we begin with the Hanning smoothed data covering the velocity range from -450 to 450 km s$^{-1}$.  As in \cite{Levine2006}, we apply a median filter to the LAB data to remove small scale features, as we are not interested in these for building our surface density maps.  This filter effectively removes small angular features from the LAB data, since we are interested in the global Galactic structure rather than the exact structure of sub-cloud sized pockets of ISM.
To construct the median filter, we cycle through each point in the LAB data and replace it with a median of itself and the nearest twelve points in the $\ell$-b plane. These 12 points will be at the same velocity, and form a diamond shape in the $\ell$-b plane centered on the point in question. Any NAN's in the \hl{brightness} temperature array, as well as any values less than or equal to 0, will be replaced by 0.01 K \hl{for the purposes of the median filtering only. This ensures that small features with large brightness temperatures that get replaced during the median filtering are not replaced by a NAN or a negative brightness temperature.}

The final piece of processing is simply to remove all \hl{brightness} temperatures with $|b| > 30^{\circ}$, which corresponds to material off of the MW disk plane, and may include features that should be excluded from our maps. The resulting \hl{brightness} temperature grid is the one used in the analysis in the main paper. \hl{The same procedure is followed for the HI4PI data for consistency.}

\begin{figure*}
\begin{center}
\includegraphics[width=\textwidth]{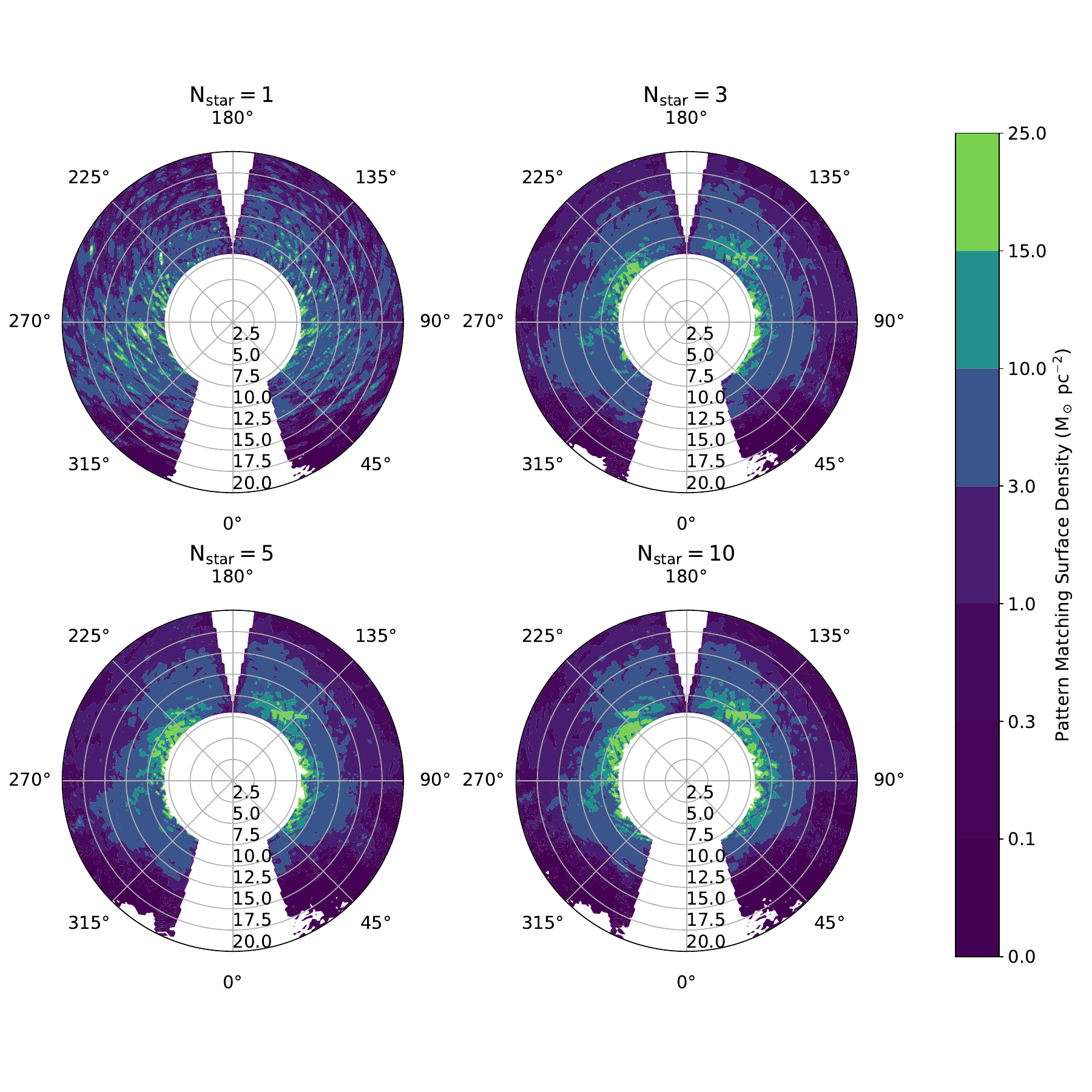}
\caption{The four panels in this figure display surface density maps generated with the pattern matching algorithm. Each map assumes a different value for $N_{\mathrm{star}}$, ranging from 1 to 10. The map labelled 3 is identical the the surface density map presented in the main paper, for comparison purposes. While there are some small scale differences in the map, the impact in the inner disk (which is most relevant for the identified spiral structure) is minimal.} \label{fig:Nstar_Panels}
\end{center}
\end{figure*}

The pattern matching algorithm relies on choices of a few parameters, and we vary these parameters here in turn, beginning with the number of stars used for distance interpolation, $N_{\mathrm{star}}$. We have varied $N_{\mathrm{star}}$ from 1 to 10, and found that the effects on the resulting map are typically small enough that they will not impact the results of this work. The exception is the $N_{\mathrm{star}} = 1$ case, which is likely unreliable as distances for the gas is determined by a single star. 
The dependence of the surface density on the choice of $N_{\mathrm{star}}$ can be seen in Figure \ref{fig:Nstar_Panels}. 


\begin{figure*}
\begin{center}
\includegraphics[width=\textwidth]{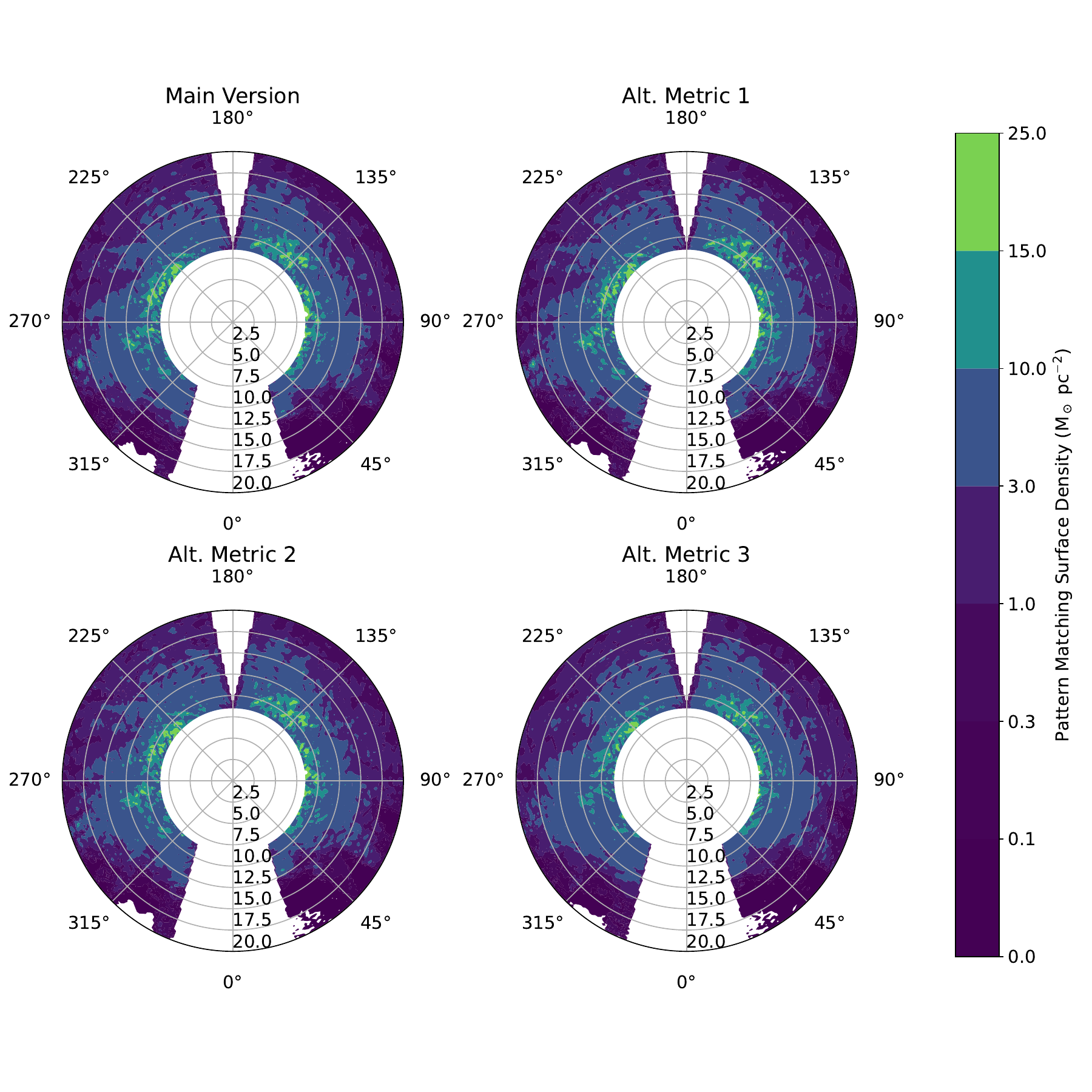}
\caption{Surface density maps generated using different distance metrics. The top left panel is the same map presented in the main paper. The top right panel was generated using Equation \ref{eqn:alt1}. The bottom left and bottom right panels were generated form Equations \ref{eqn:alt2} and \ref{eqn:alt2}, respectively.} \label{fig:dmetric}
\end{center}
\end{figure*}

Additionally, we have varied the selection of the distance metric used to match the gas measurements to nearby stars. Figure \ref{fig:dmetric} shows the results of this analysis, with the surface density maps resulting from our original selection as well as three alternative distance metrics. Each of these alternative metrics produces a similar resulting map, with differences tending to be small in scale. For a typical gas measurement, this might adjust the weighting in the weighted average, and in some cases change the set of stars from which the distance is measured.

Alternate metric \#1 changes the scaling of the velocity term, and is given in Equation \ref{eqn:alt1}. Alternate metric \#2 adds coefficients of 10 in front of the terms for the latitude and longitude, effectively producing larger weights for the position compared to the velocity. This is given in Equation \ref{eqn:alt2}. Alternate metric \#3 adds a coefficient of 10 for the velocity term, as shown in Equation \ref{eqn:alt3}.

\begin{equation}\label{eqn:alt1}
s_{LBV} = \sqrt{\frac{ (\ell_1 - \ell_2)^2 }{1^{\circ 2}} + \frac{(b_1 - b_2)^2}{{1^{\circ 2}}} + \frac{|(v_1 - v_2)|}{(1 \mathrm{km\ s^{-1}})}}
\end{equation}

\begin{equation}\label{eqn:alt2}
s_{LBV} = \sqrt{10 \frac{ (\ell_1 - \ell_2)^2 }{1^{\circ 2}} + 10 \frac{(b_1 - b_2)^2}{{1^{\circ 2}}} + \frac{(v_1 - v_2)^2}{(1 \mathrm{km\ s^{-1}})^2}}
\end{equation}

\begin{equation}\label{eqn:alt3}
s_{LBV} = \sqrt{\frac{ (\ell_1 - \ell_2)^2 }{1^{\circ 2}} + \frac{(b_1 - b_2)^2}{{1^{\circ 2}}} + 10\frac{(v_1 - v_2)^2}{(1 \mathrm{km\ s^{-1}})^2}}
\end{equation}

If the stellar density is high, which is especially true in the solar neighborhood, then the stars available are close matches in all three of coordinates, so the effect of altering the distance metric is largely negligible. This typically corresponds to regions at lower distances, where the gas density is larger (and more relevant for searching for structure in the disk). When the stellar density is low, the number of nearby stars is limited, and the effects of changing the distance metric are likely to be more pronounced. For most of the extent of our map, these regions align with the lower gas densities in the outer disk, which have limited impact on our final conclusions. 

\begin{figure*}
\begin{center}
\includegraphics[width=\textwidth]{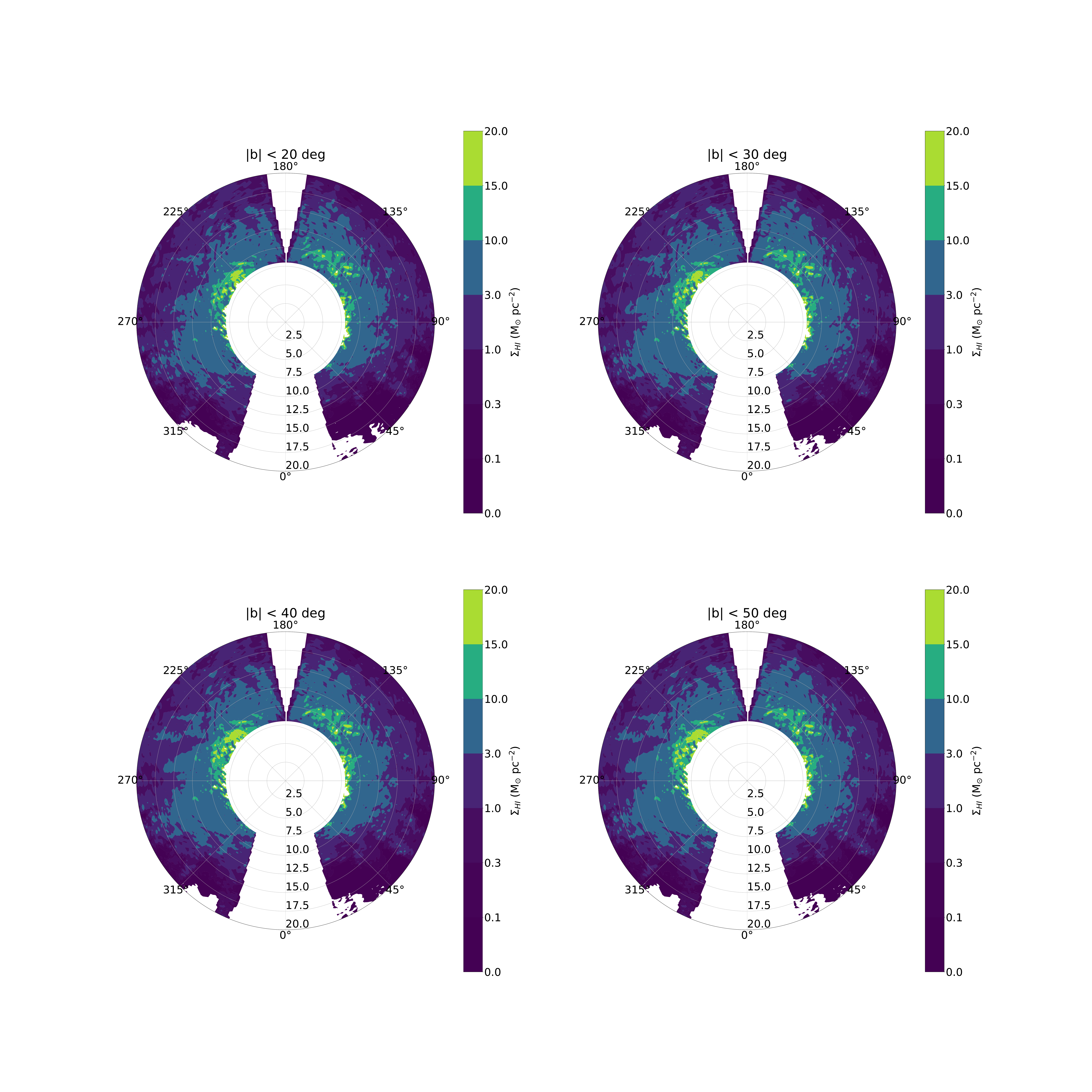}
\caption{\hl{Maps of the derived surface density for the LAB data with four different selections for the Galactic latitude cut on our data. The upper right panel selects a limit of 30 degrees off of the plane, which is the selection made across our main results. The differences that result from altering the assumed latitude cutoff are generally small.}} \label{fig:Sig_bcuts}
\end{center}
\end{figure*}

\hl{To check for the effects of the selected cut on the Galactic latitude, we have repeated our analysis for alternative latitude cuts of 20, 40, and 50 degrees. The impact on our surface density maps is quite minimal, and for the most part, there are not significant differences in the derived disk thickness as a result of these altered latitude cuts. There is one new region of increased disk thickness that is identified at larger selections for the latitude cut, which occurs at radii near 18 kpc at a $\phi$ position of 250 degrees. This may be seen below in Figure \ref{fig:th_bcuts}. In addition to the disk thickness maps, we have also checked the surface density maps and find that there are no differences in the large scale features of the disk that result from changing the cut on latitude. We especially note that the surface density in the solar neighborhood, where these cuts have the strongest limitation on the height off of the disk, the impact on the surface densities are negligible. Most of the noticeable changes in the resulting surface density maps are a result of small increases in the surface densities at some locations beyond 15 kpc in radius. These surface density maps are displayed in Figure \ref{fig:Sig_bcuts}.}

\begin{figure*}
\begin{center}
\includegraphics[width=\textwidth]{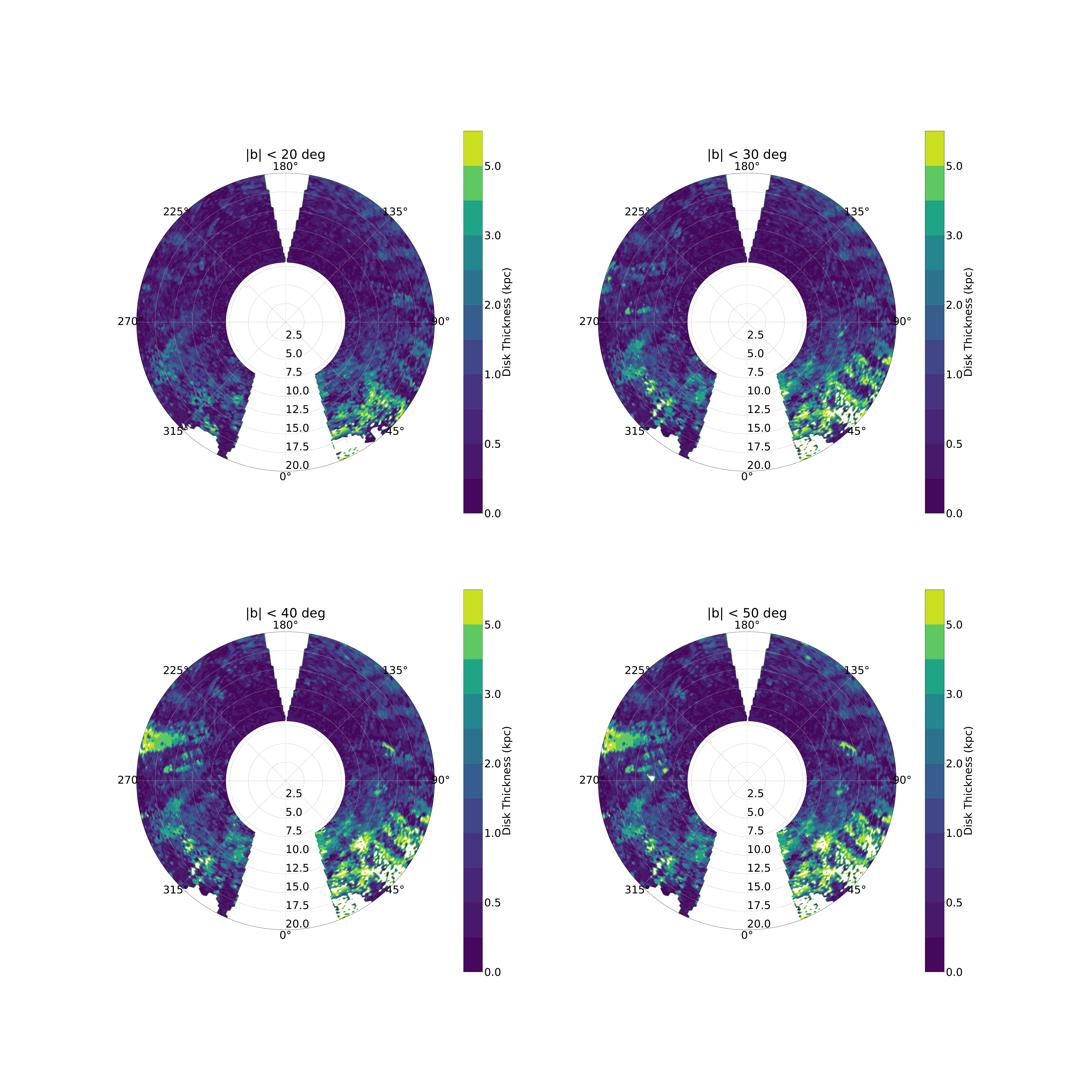}
\caption{\hl{Maps of the derived disk thickness for the LAB data with four different selections for the Galactic latitude cut on our data. The upper right panel selects a limit of 30 degrees off of the plane, which is the selection made across our main results. We can see in the bottom row, with higher latitude cuts, that there is a new region of increased disk thickness that is evident at the outer edge of this map near $\phi$ = 250 degrees. The large scale features of the disk thickness map remain the similar as this cut is altered however.}} \label{fig:th_bcuts}
\end{center}
\end{figure*}

\section{Likelihood Analysis}\label{app:likelihoods}

To test how accurately the pattern matching HI map traces the distribution of gas in the disk, we determine the likelihood that the stars and the gas are drawn from the same distribution. \hl{At a given point in longitude-velocity space, this function is defined as the nearest brightness temperature in the \HI{} data,  divided by the sum of all available brightness temperatures. To define a likelihood for a given stellar sample, we evaluate this function at the location of each star in the sample and take the product across the entire stellar sample.}

The likelihoods calculated for the stellar population are then compared to the likelihoods for randomly drawn samples using a likelihood ratio. Our likelihood ratios are calculated based on the log likelihoods, so here all likelihood ratios are given as a difference between two log likelihoods. We do this across stellar samples of different size and age, where for a given size we randomly select a subset of our stellar population to match the desired size. If the likelihood for the stars is significantly larger than the likelihood from the random sample, then the stars are tracing the gas distribution to some extent. Our randomly drawn samples start with a uniform distribution across the entire space (in \lbv{}) covered by the gas. This distribution is not a fair comparison however, as this includes random points drawn in regions that would correspond to points far outside the disk, where there is no gas (and not many stars). If the stellar population does not trace the gas but is mostly confined to the disk, it would still have a much larger likelihood than this distribution.

\begin{figure*}
\begin{center}
\includegraphics[width=\textwidth]{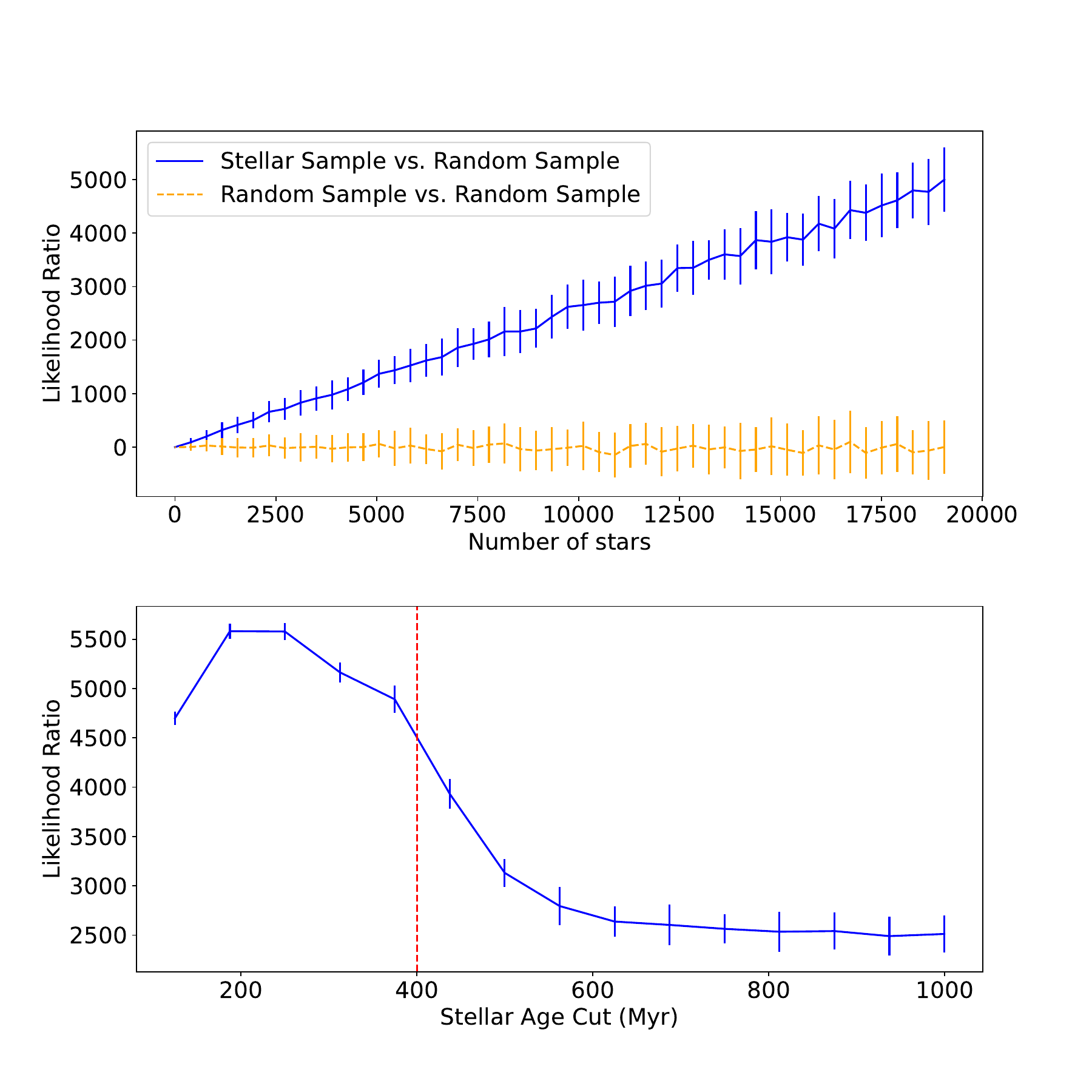}
\caption{The top panel displays the likelihood ratio between random samples of different sizes and the true stellar sample. The solid blue line displays the main results, comparing the real stellar sample against randomly drawn samples. Large ratios indicate that the true sample is more likely to be drawn from the same distribution as the gas than the random sample is. In the bottom panel, the likelihood ratios are plotted against the upper age limit of the stellar sample.} \label{fig:likelihood}
\end{center}
\end{figure*}

To remedy this, we then apply rejection sampling to reject sources below a given log likelihood, chosen to be at $ln(\mathcal{L}) = 5 \times 10^{-9}$. This particular value was selected because values much higher than this begin encroaching on the bulk of the gas distribution, while lower values begin including small pockets far away from the disk.  We randomly draw replacements until we have a sample of the desired size, all of which are above the log likelihood threshold. \hl{Lower values for this cutoff do lead to a better match with the gas, so all of our results have been tested with cutoffs extending up to $ln(\mathcal{L}) = 1 \times 10^{-7}$. At this point, there are gaps inside the region occupied by the disk that happen to be at low brightness temperatures, and therefore this distribution is expected to trace the gas a bit better than a purely random selection.} This sample will lie in the same region as the majority of the gas, but will not strongly trace any features in the gas. We then vary the size of our samples, both by randomly drawing a subset of the stars and by producing these random samples. When the likelihoods for these are computed, we get a much higher likelihood for the stars than we do for any of the random samples, which holds even for stellar samples much smaller than our real sample. These likelihood ratios, \hl{for our primary likelihood cutoff,} can be seen in Figure \ref{fig:likelihood}. \hl{We note that the slope seen in the top panel scales with the likelihood cutoff, but at the larger sample size end, the stellar sample outperforms the random sample across the full range of reasonable cutoffs.}

In this figure, the solid blue line shows our likelihood ratios across a range of sample sizes. The orange dashed line compares two randomly drawn samples to each other, to give a sense of what the likelihood rations might look like if the stellar sample was randomly distributed across the disk. The large likelihood ratios that have been observed here indicate that the actual stellar sample traces the gas significantly better than a random distribution of stars. In particular, the likelihood ratio is much larger than what is expected if the stars are not correlated with the gas. \hl{Uncertainties are estimated by measuring the likelihood ratios across 50 different randomly drawn samples, and taking the standard deviation of the derived likelihood ratios.}

There is an age dependence on the strength of this correlation, and younger stars provide better likelihood ratios than older stars, as expected. An entirely arbitrary selection of Gaia sources, established with no age cuts and designed to select the same size stellar population as our actual sample, is a good point of comparison. In this case we no longer trace the gas better than a randomly drawn sample, so old stars selected from Gaia are not a good selection for this case. Shown in figure \ref{fig:likelihood} is the likelihood ratio as a function of the age cut on our stellar sample. At lower age cuts we obtain better likelihood ratios, except at very low ages where the ratio drops again, likely due to significantly reduced sample sizes. \hl{We note that at the large age end, where we are close to selecting a random set of Gaia sources, it remains the case that the Gaia sample traces the underlying gas distribution to some extent. The best match is obtained from stars younger than 400 Myr, and the likelihood ratio flattens out at around 600 Myr.}

\section{Additional Simulations}\label{app:sims}

\begin{figure*}
\begin{center}
\includegraphics[width=\textwidth]{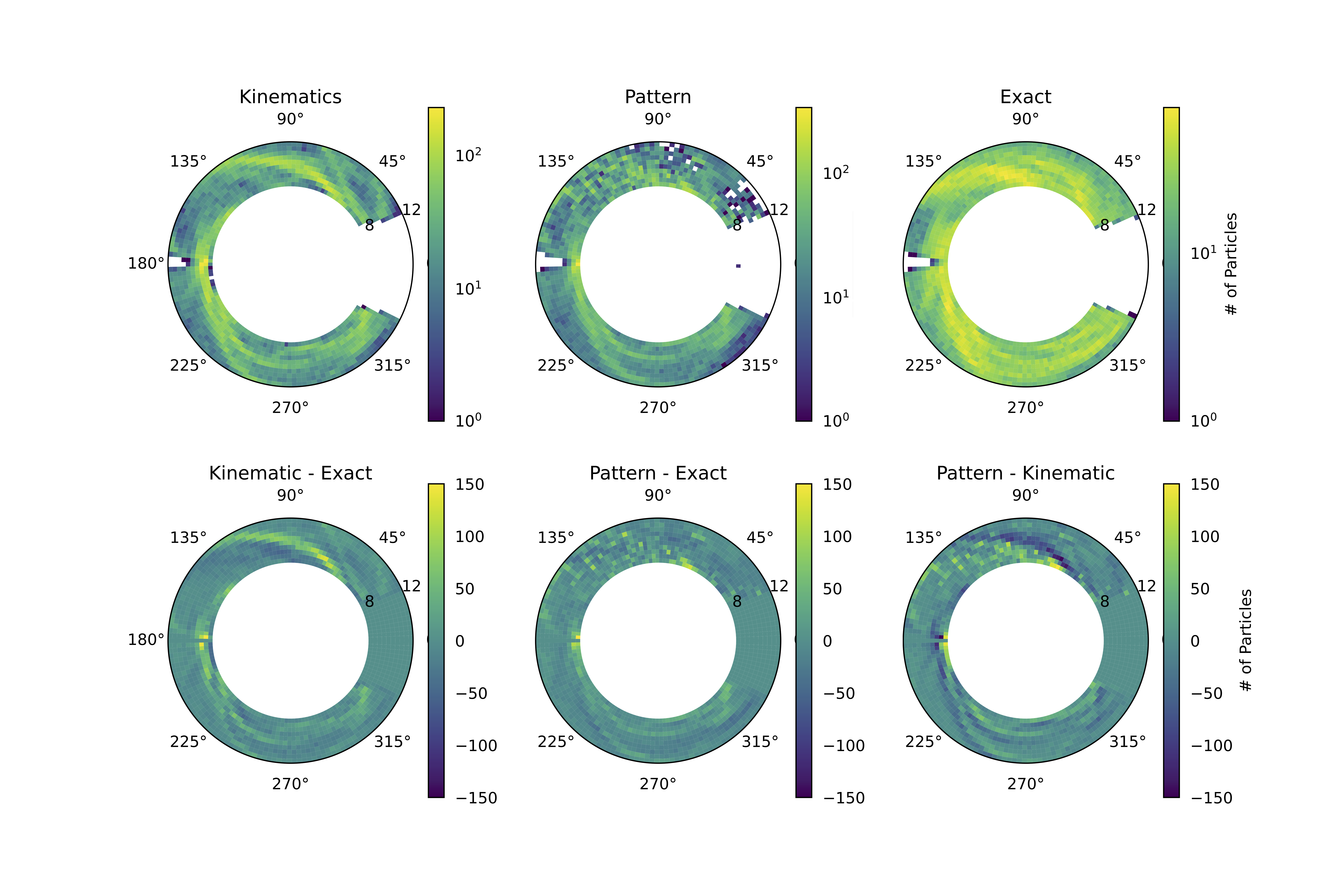}
\caption{Face on gas distributions from a snapshot of the Sagittarius simulation. Colorbars in each panel display the number of particles in each bin. The left column shows results from kinematic estimates, with the resulting HI distribution in the top panel and the residual shown in the bottom panel. The middle column shows the results with the pattern matching distances. The top right panel shows the exact HI distribution from the simulation. The bottom right panel shows the difference between the maps derived using the pattern matching and kinematic distances.} \label{fig:SGR_Faceon}
\end{center}
\end{figure*}

In addition to the SP2 simulation discussed in the main paper, we have also analyzed a more dynamically complex simulation that contains three dwarf galaxies orbiting a MW-like galaxy. The three orbiting satellites are the two Magellanic clouds and the Sagittarius dwarf galaxy. This simulation will be referred to as the SGR simulation. We have evaluated many snapshots throughout this simulation, and the pattern matching technique typically provides a better match to the structures in the simulated disk than the kinematic map does. \hl{Maps from the present-day snapshot of this simulation are shown in Figure \ref{fig:SGR_Faceon}.}

In this particular simulation, the range of radii for which the pattern matching technique is viable is somewhat limited, as a result of the lack of newly formed star particles at radii greater than 12 kpc. This, combined with the limitations for gas inside the solar circle, results in a narrow range of radii for which it is possible to generate a surface density map. For comparison purposes, we limit our analysis to the range of radii for which there is a sufficient stellar population to utilize the pattern matching technique.

We have also analyzed a simulation that is similar to the SP2 case but with 4 spiral arms. For testing purposes the SP2 case is simpler to analyze due to the reduced number of arms, and it makes a good test case for our algorithms. Additional simulations include an isolated disk and a case with a dwarf galaxy on a coplanar orbit. All of these produce similar accuracy results to those presented in SP2, so we adopt SP2 as our fiducial test model.

\end{document}